\documentclass[twocolumn,pra]{revtex4-1}
\pdfoutput=1

\usepackage{xspace}
\usepackage{pifont}
\usepackage[pdftex]{graphicx}
\usepackage[utf8]{inputenc}
\usepackage[T1]{fontenc}
\usepackage{graphics}
\usepackage{graphicx}
\usepackage{textcomp}
\usepackage{amsmath,amssymb}
\usepackage[english]{babel}
\usepackage{xspace}
\usepackage[pdftex,hidelinks]{hyperref}
\usepackage{xcolor}

\newcommand{\D}{$\Delta$n\xspace}

\begin{document}

\graphicspath{{./},{./Figures/}}

\title{Ultrafast Laser Inscription of High Performance Mid-Infrared Waveguides in
  Chalcogenide Glass}

\author{Pascal~Masselin}\email{masselin@univ-littoral.fr}
\affiliation{Université du
  Littoral-Côte d'Opale, Lab. Physico-Chimie de l'Atmosphère, F-59140
  Dunkerque, France}%
\author{Eugène~Bychkov}
\affiliation{Université du
  Littoral-Côte d'Opale, Lab. Physico-Chimie de l'Atmosphère, F-59140
  Dunkerque, France}%
\author{David~Le~Coq}
\affiliation{Université de Rennes, CNRS, ISCR - UMR 6226,
  F-35000 Rennes, France}%

\begin{abstract}
  We present the realization of mid-infrared waveguide by ultrafast
  laser inscription technique in
  a chalcogenide glass. Our approach is based on multicore waveguide
  that consists in an alignment on a mesh of positive refractive index
  channels placed parallel to each other. Two different meshes are
  investigated with different refractive index contrasts between the
  channel and the glass matrix. A detailed analysis of the
  performances at a wavelength of 4.5 µm shows propagation
  losses of 0.20 $\pm$ 0.05 dB/cm and coupling efficiencies higher than 60\%.
\end{abstract}

\maketitle

\section{Introduction}

In the mid-infrared (mid-IR) spectral range, between 2 and 25 µm, most
of the molecules have strong absorption peaks that can be used to
selectively and efficiently detect and quantify chemical or biological
species.  This is the reason why large-scale projects are currently
developing around mid-infrared photonics for various applications
directly concerning important societal issues like health and
environment monitoring or security \cite{MIRTHE,Minerva,FLAIR,MIRIF}.

The extension of silicon technology to higher wavelengths is a natural
way to make mid-IR photonic components.  However silicon oxide that is
used in this technology, rapidly limits the possible wavelength range
since it starts to absorb around $\sim$3.6 µm
\cite{Soref2010}. Therefore technical solutions to minimize the amount
of light propagating in the SiO$_2$ layer should be developed.  A
first approach is to modify the geometry of the guides to minimize
contact between the core and the absorbing layer, for example, by
suspending them \cite{Penades2016} or placing them on a pedestal
\cite{Lin2013}.  The obtained propagation losses are below 1 dB/cm
(0.82 dB/cm precisely) in the case of suspended waveguide and 2.7
dB/cm in the case of pedestal use but both of these results are
obtained for a wavelength below 4 µm. An other strategy more suitable
for higher wavelengths is to modify the material. For an example
silicon on sapphire nanowires have shown propagation losses of 2 dB/cm
at $\lambda = 5.18$ µm \cite{Li2011} and 1 dB/cm at $\lambda = 4$ µm
\cite{Singh2015}.  On the other hand germanium-based materials seem to
be very promising especially for high wavelengths although currently
the level of propagation losses remains above 1 dB/cm
\cite{Chang2012,Brun2014,Ramirez2017,Ramirez2018}. Moreover the small
physical dimensions of the transverse section of those waveguide
preclude efficient light collection and coupling losses of several
decibels are often reported.

Another way to obtain a waveguide beyond 4 µm is to make it by ultrafast
laser inscription (ULI) technique in glass. The ULI is a very versatile
and cost-effective method for rapid prototyping and production of
optical components \cite{Osellame2012}. It has been applied in the mid-IR to
different chalcogenide glasses
(gallium lanthanum sulfide and
75GeS$_2$-15Ga$_2$S$_3$-4CsI-2Sb$_2$S$_3$-4SnS in \cite{Rodenas2012}
and Ge$_{15}$As$_{15}$S$_{70}$ in \cite{DAmico2014})
and guiding up to $\lambda = 10$ µm has been experimentally
demonstrated. Very recently propagation loss around 1-1.5 dB/cm has
been reported at $\lambda = 7.8$ µm in an other germanium based
composition (Ge$_{33}$As$_{12}$Se$_{55}$) \cite{Butcher2018}.

We have previously reported a writing procedure of multicore waveguide
in an arsenic-free germanium based chalcogenide glass
(72GeS$_2$-18Ga$_2$S$_3$-10CsCl) that shows propagation loss of 0.11
$\pm$ 0.03 dB/cm at $\lambda = 1.55$ µm \cite{Masselin2016}. In this
letter we presents the extension of these results to the mid-IR at
$\lambda = 4.5$ µm.

\section{Description of the writing procedure}

The photowritten waveguides are multicore type and they consist of channels
of positive refractive index variation (\D) induced by a train of
femtosecond pulses, placed parallel to each other on a 
mesh. Here the inscription geometry differs from the classical
transverse or longitudinal ones as the irradiation is done without
continuous sample translation. In a first step the laser beam is
focused in front of the channel \ding{172} and the sample is
irradiated with a burst of femtosecond pulses. The duration $\tau$ of
this burst is an important parameter of the experiment as it will be
described later. The result of this irradiation is an increase of the
length of the channel \ding{172}. In a second step the sample is
translated perpendicularly to the writing direction in the plan of the
transverse section so that the channel \ding{173} is in front of the
focus of the laser beam. A second burst of pulses is sent over the
sample which again increases the channel length. The operation is
repeated as needed for all the channels until the slice of transverse
section is completed. Then the sample is translated parallel to the
writing beam and the procedure is repeated over the entire length of
the sample.

\begin{figure}[htbp]\centering
  \includegraphics[width=\linewidth]{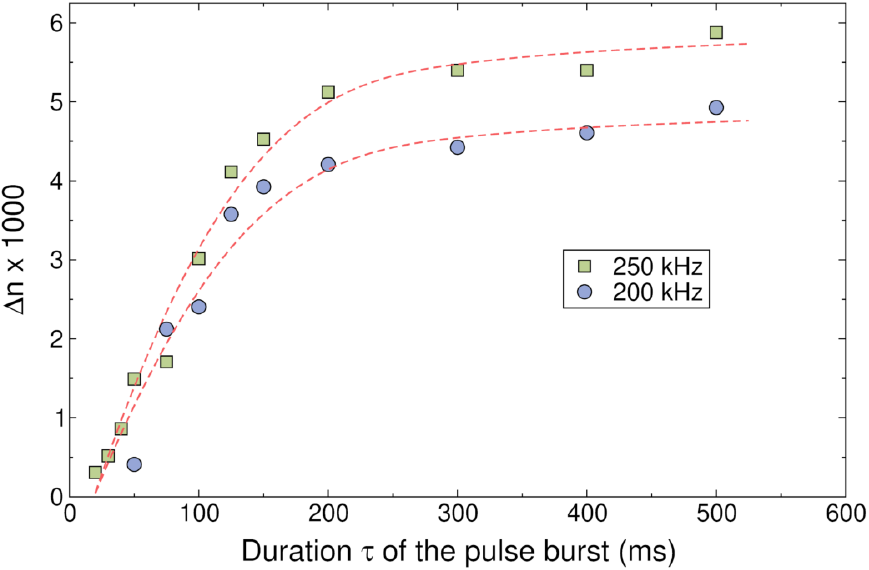}
  \caption{Dependency of the refractive index contrast between the
    individual channel and the glass matrix for different repetition
    rate of the pulse train. The dashed lines are a guide for eye.}
  \label{Fig:Dn_vs_Tau}
\end{figure}

This procedure presents the advantage that the magnitude of the
refractive index contrast between the channel and the non-irradiated
matrix can be easily controlled by varying the duration of the burst
$\tau$. We have measured \D using quantitative phase microscopy
followed by an Abel inversion \cite{Ampem-Lassen2005} and its
dependency with $\tau$ is presented on the figure \ref{Fig:Dn_vs_Tau}
for different repetition rates of the pulse train. It can be seen from
this figure that \D increases nearly
linearly for $\tau$ values below 150 ms and saturates for higher
values. Also the level of saturation depends on the repetition rate,
which can be attributed to the phenomenon of accumulation of charges
released during  the writing process \cite{Caulier2011}.

On the other hand, the diameter of the individual channels is constant
due to the formation process of \D. Indeed, the refractive index
variation is related to the formation of a filament during the
propagation of the femtosecond pulse in the glass, whose diameter is
defined by the properties of the material\cite{Caulier2011}. The
thermal effects that could induce a dependence of the diameter of the
inscription with the experimental parameters \cite{Caulier2013}, are
not dominant in the process of appearance of \D.  However it is
possible to adjust the diameter of the total structure by changing the
distance between the channels or by adding others. Therefore it is
possible to select independently \D and the dimension of the waveguide
in order to engineer its characteristics as needed.

\section{Waveguide performance measurements}

\begin{figure}[htbp]\centering
  \includegraphics[width=\linewidth]{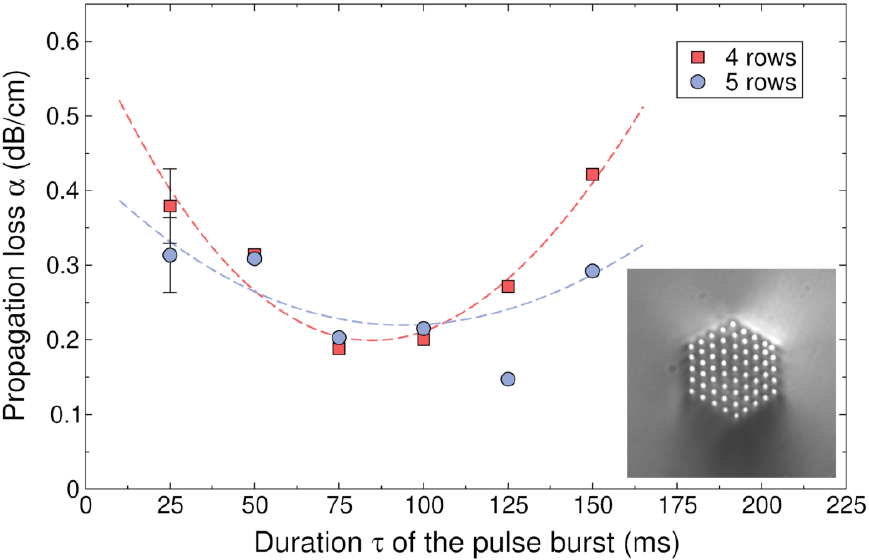}
  \caption{Measured propagation losses for hexagonal mesh structures
    with 4 and 5 rows. The dashed lines are a guide for eye.}
  \label{Fig:Loss-Hexa}
\end{figure}

First we study the case of waveguide composed of channels placed on a hexagonal
mesh and two configurations are compared. They differ in the number of
rows that make up the structure. The first is composed of 4 rows of
channels separated by a distance of 2.875 µm and the second by 5 rows
with a separation of 2.3 µm between the channels so that the total
diameter of the structure is the same (23 µm). A typical example of
such structure is represented on the inset of the figure
\ref{Fig:Loss-Hexa}. The waveguides are written with a laser repetition rate
of 200 kHz.

The propagation losses are measured according to the back reflection
method \cite{Ramponi2002}. After the inscription the sample is cut and
repolished to eliminate edge effects. The input face is cut at a
slight angle to the cross section plane of the guides in order to
spatially separate the reflected waves on the input and the output
face of the waveguide and thus eliminate interference likely to
distort measurements. Their total length after this step is equal to
28 mm. The results of the loss measurement are shown in the figure
\ref{Fig:Loss-Hexa}.  The existence of an optimal irradiation duration
value $\tau_{min}$ that leads to a minimum of propagation losses
$\alpha$ is clearly observed. This minimum value $(\alpha_{min})$ is
equal to 0.20 $\pm$ 0.05 dB/cm, which is far below the values obtained
for photowritten waveguides or those derived from silicon photonics in
the mid-IR. In fact for $\tau < \tau_{min}$ the beam is poorly
confined in the structure whereas for $\tau > \tau_{min}$ the field
tends to be localized into the individual channel. The optimal
condition is a compromise between these two situations
\cite{Masselin2016}.

It is interesting to note that $\alpha_{min}$ is the same for the two
configurations (4 and 5 rows). This means that it is independent of
the density of individual channels, i.e. the number of channels per
square micron, since the surface of the transverse section is the same
for both configurations. Of course it will not be the case if the
distance between the channels becomes too large (i.e. structure with 2
or 3 rows, considering the same total diameter of written
structure). Therefore it is a good indication that our writing method
leads to very homogeneous refractive index variation channel and that
the concatenation of the different slices of transverse section does
not add irregularities that would scatter the light similarly to side
roughness.  However for a 4 rows structure with a lower density (0.177
channels/µm$^2$ compared with 0.265 channels/µm$^2$ in the case of 5
rows) the optimal value of $\tau$ becomes more critical since the
dependence of $\alpha$ over the burst duration $\tau$ is more
pronounced.

It should be noted that all of these guides, and those described later
in the text, were written in the same piece of glass. Thus the properties of
the host matrix, which could influence the performance, are the
same for all waveguides and the differences between the behaviors of the
propagation losses are due only to the different structures of the
transverse section.

\begin{figure}[htbp]\centering
  \includegraphics[width=\linewidth]{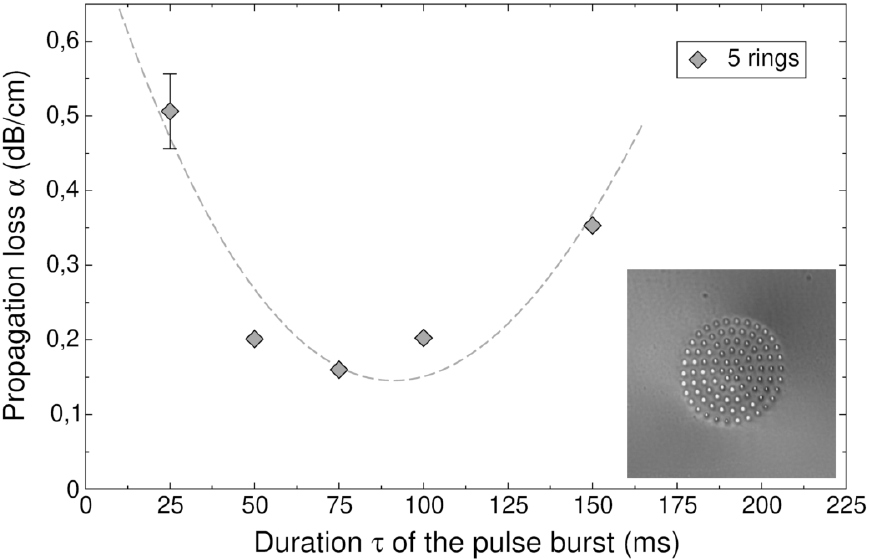}
  \caption{Measured propagation loss for circular mesh structure with
    5 rings. The dashed line is a guide for eye.}
  \label{Fig:Loss-Circ}
\end{figure}

To illustrate the versatility of our method we modify the morphology
of the transverse section of the waveguide. Here the mesh is composed
of concentric rings. On the N$^{\text{th}}$ ring the channels are
separated by an angle $2\pi /6N$. Here the inscriptions were made with
a laser repetition rate of 250 kHz. Indeed, preliminary evaluations
carried out with a rate identical to that used for the hexagonal mesh,
have shown that higher values of \D are necessary to obtain minimum
losses. A photograph of a structure formed of 5 rings is shown in the
inset of the figure \ref{Fig:Loss-Circ} that presents the propagation
losses for different burst durations.  On this figure we can see that
the behavior of $\alpha$ with $\tau$ is the same as for a hexagonal
mesh and that the minimum value is the same within the uncertainties
of the measurements. Here also, we can observe that the decrease in
the \D channels density (0.219 channels/µm$^2$) leads to a stricter
dependence of propagation losses with the duration of the pulse burst
compared with the hexagonal mesh with 5 rows.

We can now consider the other sources of losses but before we note
that the intrinsic absorption of the bulk material is already taken
into account in the measurement of the propagation losses. First of all the
reflection coefficient at the air-guide interfaces was measured as
being in the order of 13\%, which is in agreement with the value of the
refractive index of the sample \cite{Masselin2012}.
We also measured the power carried by the waveguide and derived the
coupling efficiency using the following equation:

\begin{equation}
  \eta = \frac{1}{\left( 1 - R \right)^2 \, T_{Opt}} \, 10^{\left(0.1 \,
    \alpha  L\right)} \, \frac{P_{Out}}{P_{In}}
\end{equation}
where $R$ is the reflection coefficient from the interface between air
and the waveguide, $T_{Opt}$ is the transmission coefficient of the
focusing and collimating optics, $\alpha$ is taken from the values
reported in the figures \ref{Fig:Loss-Hexa} and \ref{Fig:Loss-Circ},
$L$ is the length of the waveguide (28 mm) and $P_{In}$ and $P_{Out}$ are the
powers measured before and after the waveguide, respectively.

\begin{figure}[htbp]\centering
  \includegraphics[width=\linewidth]{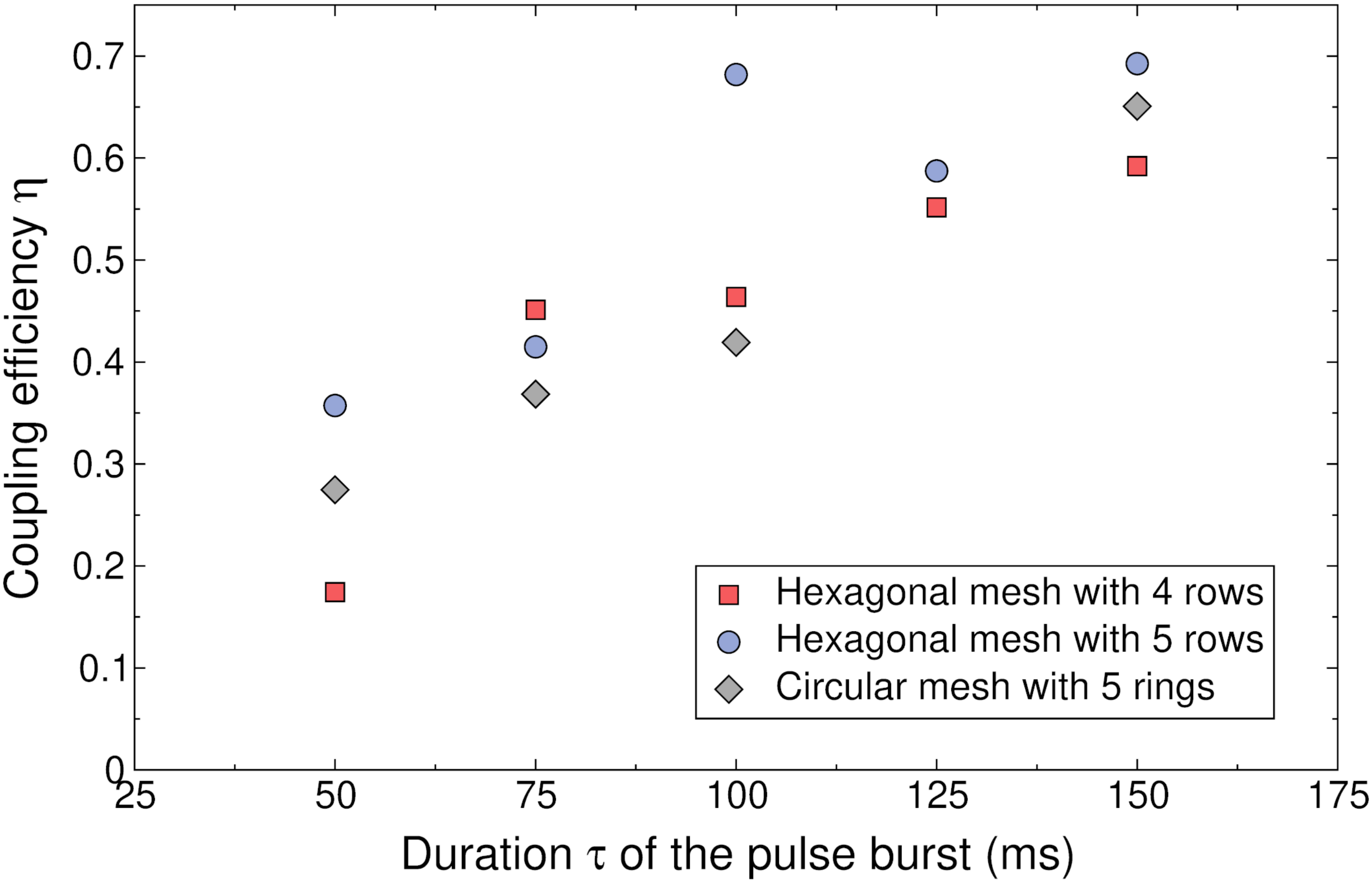}
  \caption{Coupling efficiency of light inside the waveguides for the
    hexagonal and circular meshes.}
  \label{Fig:Coupling}
\end{figure}

The results of these measurements are reported on the figure
\ref{Fig:Coupling}. For all the considered structures the coupling
efficiency $\eta$ is an increasing function of the burst duration,
i.e. \D, with values that can be higher than 0.6. However if the
behavior of the dependence is the same for all meshes, we note that
for most of the values of $\tau$, 
the highest value for $\eta$ is obtained for the 5 rows hexagonal mesh
that corresponds to the structures with the highest density of \D channels.

\begin{figure}[htbp]\centering
  \includegraphics[width=\linewidth]{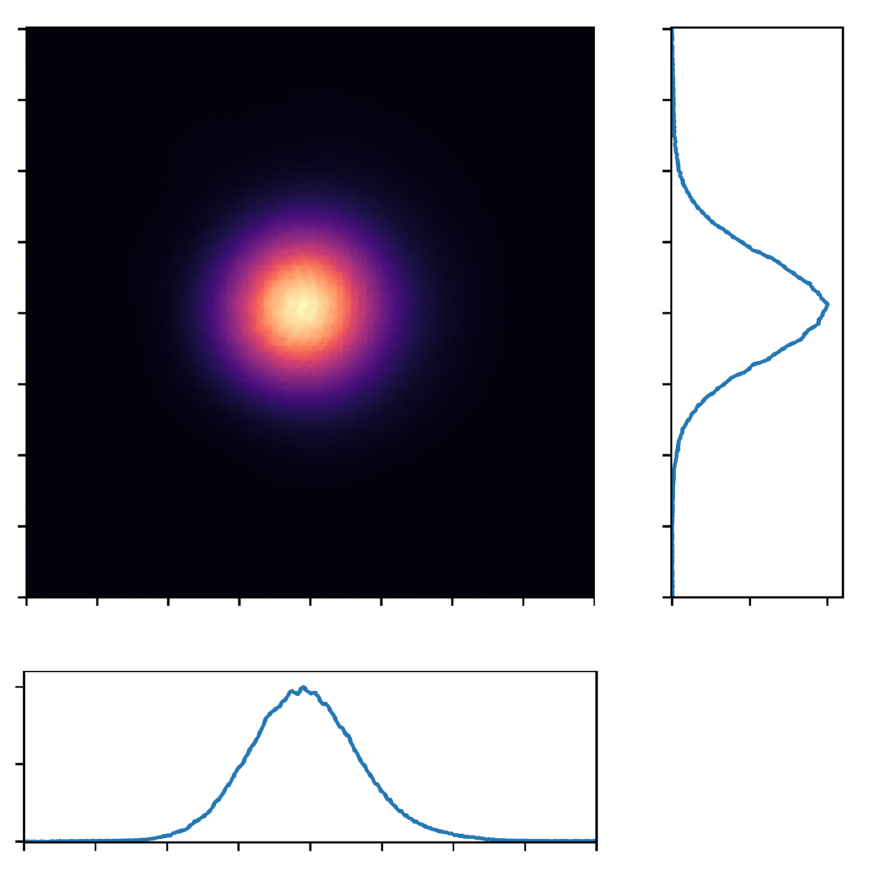}
  \caption{Mode of the propagated beam in a hexagonal mesh with 5
    rows. Similar profiles are obtained with 4 rows and with circular mesh. }
  \label{Fig:Mode}
\end{figure}

It should be remarked that the global performance of the waveguide
being a combination of both the propagation loss and the coupling
efficiency, the value of $\tau$ that gives the lowest value
of $\alpha$ could be not the optimum one since a higher value would
increase $\eta$ and at last, leads to a maximized carried power.
Consequently $\tau$ should be chosen according to the functionality
of the device. For a short one it would be preferable to select large
$\tau$ to maximize $\eta$ while if the length is so that the
propagation losses are predominant, $\tau$ should be shorter to minimize $\alpha$.

Finally the near field pattern of the propagated beam was imaged on
the InSb sensor of a mid-IR camera (FLIR A6750sc) and the mode has
been measured for the optimum values of the burst duration $\tau$. As
it can be seen from the figure \ref{Fig:Mode} the beam profile is very
close to be Gaussian indicating the single mode behavior of the
waveguide. On this figure one can see the mode corresponding to a
hexagonal mesh with 5 rows but similar profiles are obtained for the
others structures.

\section{Conclusions}

In conclusion we present the realization of mid-IR waveguides in
arsenic free chalcogenide glass with propagation loss below 0.2 dB/cm
at 4.5 µm. Moreover our method allows a control of the waveguide
dimensions meaning that an efficient light collection can be achieved
in most situations as the diameter of the structure can be adapted
according to the requirement, i.e. butt coupling from a fiber or
coupling from free space. As an illustration we are able to achieve in
the best configuration a total power carried corrected for Fresnel
losses is as large as 60\% of the incident power (we do not take into
account Fresnel loss since it can be minimized by using an
anti-reflective coating).  By combining these results with those
obtained previously \cite{Masselin2016} this demonstrates that the
described method can be used to design waveguide devices with high
performances over the whole range between 1.5 and 4.5 µm. It can be
objectively assumed that this range can be extended to higher
wavelengths and that the ultimate limit would be defined by the
transmission of glass (10 µm). This work is currently being continued
to extend this technique to the production of curved guides.

This work has been partially supported by the French National
Research Agency (ANR) through the COMI project (ANR-17-CE24-0002) and as well as
by the Ministry of Higher Education and Research, Hauts de France
council and European Regional Development Fund (ERDF) through the
Contrat de Projets Etat-Region (CPER Photonics for Society P4S).

\bibliography{Bib-OL}

\end{document}